\documentclass[11pt,twoside]{article}

\usepackage{asp2004}
\usepackage{epsf}
\usepackage{psfig}
\usepackage{lscape}
\usepackage{amsmath}
\usepackage{amssymb}

\begin{document}
\title{Present and Future Astrometric Study\\ of Halo Substructure }   
\author{Steven R. Majewski$^1$, Allyson A. Polak$^1$, David R. Law$^{1,2}$, \\
Helio J. Rocha-Pinto$^1$}   
\affil{$^1$University of Virginia, P.O. Box 3818, Charlottesville, VA 22903-0818\\
$^2$Caltech, Department of Astronomy, MS 105-24,
Pasadena, CA 91125}    

\begin{abstract} 

We focus on how astrometric surveys might be
exploited for the study of Galactic halo substructure originating
from the disruption of satellite systems.  The
Sagittarius (Sgr) dwarf spheroidal tidal stream provides a useful template to explore
 present and future capabilities of
proper motion surveys for finding and studying substructure as well as for using individual
tidal streams to probe the global properties of the Milky Way.
The configuration of the Sgr stream is particularly well suited
for measuring the flattening of the halo and for providing an
independent means for measuring the speed of the Local Standard
of Rest.  For the various proposed experiments, the required proper motion 
accuracies are well within the capabilities of 
the Space Interferometry Mission (SIM PlanetQuest).

\end{abstract}



\section{Introduction}

It is now clear the halos of spiral galaxies at least
partly originate with the accumulation of material from 
smaller
stellar systems.  The material from each accreted satellite 
inhabits defined regions of phase space to create halo substructure, often 
characterized, at least early on in the process,
by spatially coherent tidal tails.
Examples of stellar tidal streams from disrupting satellites
are now well known and well studied in the Milky Way, M31 and other disk galaxies
(e.g., Shang et al. 1998, Ferguson et al. 2002, Pohlen et al. 2004).

Majewski (1999) summarizes a variety of observational
strategies for uncovering and mapping tidal streams in the Milky Way halo.  
In general, what we know about these substructures comes from an analysis of their 
spatial and radial velocity characteristics, whereas the astrometric demands on 
the study of even relatively close
streams remains challenging, and often tantalizingly just out of reach.  This is unfortunate, because, in general, proper motion surveys
yield more dynamical information than radial velocity surveys: (1) more stars
measured at a time, and (2) two components of motion per star versus just one. 

Because of limitations in precision, 
astrometric catalogs have mainly been plied to the study of rather {\it local} halo
substructure --- primarily searches for 6-D phase space clumping among
nearby stars (e.g., Eggen 1977, 1987, 1996, and references therein; 
Poveda et al. 1992; Helmi et al. 1999). 
Majewski et al. (1996) used a very deep proper motion and radial velocity
study to push the technique to finding more distant --- though still within
10 kpc --- halo substructures, whereas
Aguilar \& Hoogerwerf (1999) describe a method for finding nearby
substructure with   
proper motion catalogs in the absence of radial velocity data.

In this contribution we explore additional applications of astrometry
to the study of 
halo substructure, both within and outside the solar neighborhood,
and look ahead to 
future possibilities with higher precision
proper motion data.

\section{The Sagittarius Stream as a Template}

As the most vivid known example of tidal disruption of a dwarf galaxy in the 
Galactic environment, the Sagittarius (Sgr) system serves as a useful guide
to the possibilities and difficulties of dynamical studies of halo substructure.
An extensive mapping of the Sgr system comes from M giants identified with
the Two Micron All-Sky Survey 
(2MASS; Majewski et al. 2003 --- MSWO here-after).  These
data show that (1) Sgr is the predominant source of M giants in the high halo, (2) such stars
are easily observed and clearly discriminated by 2MASS photometry, and (3) pieces of
the Sagittarius tidal streams come quite close to the Sun --- within reach of current
astrometric catalogs.  We adopt this
360$^{\circ}$-wrapped, Galactic polar ring of M giants as a template for our discussion
of the astrometric analysis of tidal streams.

\subsection{Models of Predicted Motions}

Law et al. (2004, 2005) have taken the available spatial (MSWO) and radial
velocity data (Majewski et al. 2004a, and in prep.) for 2MASS M giants to constrain models of the
interaction of Sgr with the Milky Way.  In general, the data are best fit by a Sgr system of
current mass 3.5 $\times 10^8$ M$_{\sun}$ and 
space velocity of 328 km s$^{-1}$ in an orbit having a period of 0.85 Gyr and an
apo:peri-Galacticon ratio of 57:13 kpc.  These models were fit to what appears to be 
approximately 2.5 orbits (or 2.0 Gyr) of Sgr mass loss in M giants. 
Additional modeling details, including
the Galactic potential used, are given by Law et al. and below (\S4). 
In brief, the Galactic potential is smooth, static and given by the sum of a 
disk, 
spheroid and 
logarithmic halo described by the axisymmetric function
\begin{equation}
        \Phi_{\rm halo}=v_{\rm halo}^2 \ln (R^{2}+[z/q]^2+d^{2}).
\label{haloeqn}
\end{equation}
where $q$ is the halo flattening, $R$ and $z$ are cylindrical coordinates and $d$ is 
a softening parameter.   

Model fits to the spatial and radial velocity 
Sgr data allow predictions of the full 6-D phase space
configuration of Sgr debris.  Figure 1 shows predictions for the $U,V,W$ velocity
components of the debris as a function of longitude in the Sgr orbital 
plane, $\Lambda_{\sun}$ (see MSWO for definition of this plane).
The debris is shown 
assuming a halo potential flattened to $q=0.9$ and assuming a characterization
of the total potential whereby the rotation curve yields a circular speed at the Sun of 
220 km s$^{-1}$.  We explore how variations in the {\it shape} of the Galactic potential
affect primarily the 
$U$ and $W$ velocities in Law et al. (2005)
and briefly in \S4, 
and how variations in the {\it scale} of the potential (i.e. the rotation curve)
affect primarily the $V$ velocities in \S2.2.

\begin{figure}[!ht]
\plotfiddle{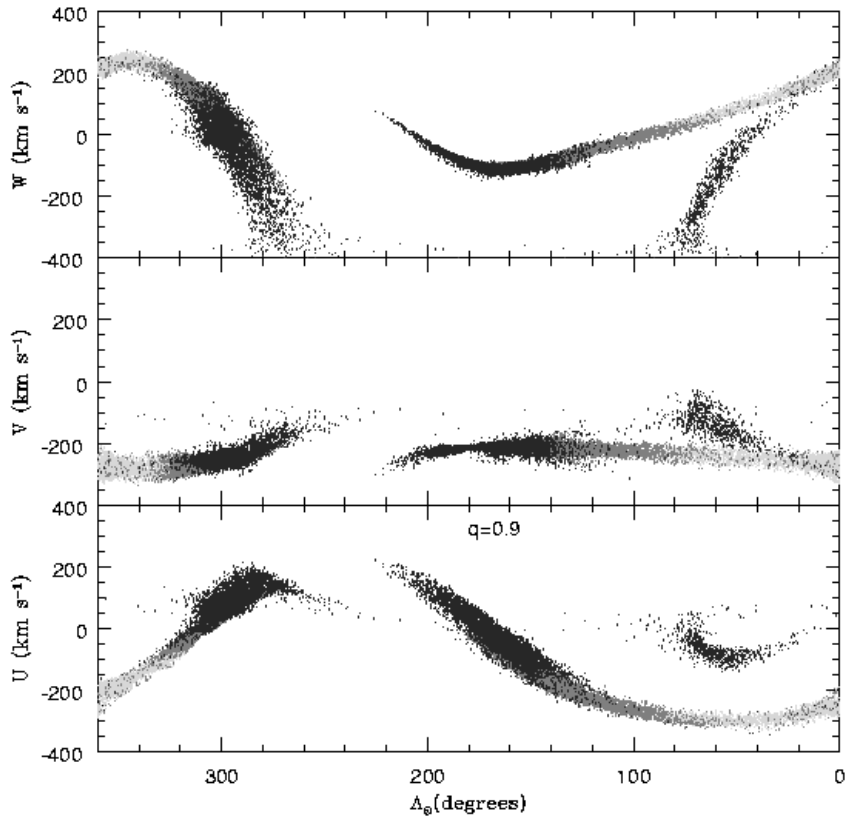}{100truemm}{0}{58}{58}{-171}{-95}
\caption{Predicted $U,V,W$ velocities as a function of longitude in the Sgr orbital plane
(with $\Lambda_{\sun}=0^{\circ}$ at the present Sgr position and increasing towards trailing debris).
Debris lost over the past 2.5 radial periods of the Sgr orbit are shown (with different greyscales
corresponding to debris lost on different apo- to apo-Galacticon orbits).  Sgr trailing debris
is on the right stretching along the continuous strand from $\Lambda_{\sun}=0^{\circ}$ to $\Lambda_{\sun} \sim
220^{\circ}$; leading arm debris starts on the left at $\Lambda_{\sun}=0^{\circ}$ and is prominent
to $\Lambda_{\sun} \sim 240^{\circ}$ at which point it crosses the Galactic plane and reappears 
to overlap in $\Lambda_{\sun}$ the trailing debris 
over $\Lambda_{\sun} \sim 140^{\circ}$ to $0^{\circ}$.  The Galactic model has a $q=0.9$ halo 
and circular
speed at the Sun of 220 km s$^{-1}$.  Variations in the $V$-velocities of Sgr trailing debris show almost imperceptible variation for adopted 
$q$ varying from 0.9 to 1.25 and adopted $R_0$ varying from 7 to 9 kpc.
}
\end{figure}

\subsection{ Sgr Debris as a Tool for Measuring $\Theta_{\rm LSR}$}

Despite decades of effort by astronomers, the rotation rate of
our Galaxy remains poorly known, with measurements of the local speed
that vary by 25\%. Hipparcos
proper motions \citep{feast97} suggest that the 
Local Standard of Rest (defined as the velocity of a closed orbit at $R_0$) is
 $\Theta_{\rm LSR}
=(217.5\pm 7.0) (R_0/8)$ km s$^{-1}$ --- i.e. near the IAU adopted value of 
220 km s$^{-1}$.
But a more recent
measurement of the proper motion of Sgr A$^*$  
(Reid \& Brunthaler 2004) yields a higher value of $(235.6\pm1.2) (R_0/8)$ km s$^{-1}$, 
whereas direct HST measurements of the proper motions of bulge stars against
background galaxies in the field of the globular cluster M4 yield $(202.4\pm20.8) (R_0/8)$ km s$^{-1}$ (Kalirai et al. 2004) and $(220.8 \pm 13.6) (R_0/8)$ km s$^{-1}$ (Bedin et al. 2003). 
Of course, these
measures (as well as any of those depending on the Oort constants) 
 rely on having an accurate measure of $R_0$; however,
with recent geometrical measurements of the motions of stars around Sgr A$^*$, $R_0$
is now known to 5\% ($7.94\pm0.42$ kpc; Eisenhauer et al. 2003).   The 
solar peculiar motion must also be known, but this is a smaller correction 
(e.g., $5.3\pm0.6$; Dehnen \& Binney 1998).
On the other hand, considerations of non-axisymmetry of the disk yield corrections to
the measurements that suggest  
 $\Theta_{\rm LSR}$ may be as low as $184 \pm 8$ km s$^{-1}$ \citep{om98} or lower
(Kuijken \& Tremaine 1994).
Because $\Theta_{\rm LSR}$ is fundamental to establishing the mass scale of the
Milky Way, independent methods to ascertain it are of great value.  

Eventually, as part of
a Key Project of the Space Interferometry Mission (SIM), both $\Theta_{\rm LSR}$ 
and $R_0$ will be measured directly and precisely
by the absolute
proper motion and parallax of stars at the Galactic center.  However, here
we describe an independent
method for ascertaining $\Theta_{\rm LSR}$ that overcomes several difficulties
associated with working in the highly dust-obscured and crowded Galactic center, 
and one that is also insensitive to $R_0$ (for all reasonable values of the latter).  
The target stars for this method ---
Sgr trailing arm stars --- are ideally placed for uncrowded field astrometry at high
Galactic latitudes, and are at relatively bright magnitudes for, and require only the most
modest precisions from, SIM.  Indeed, as we shall now show, this method is even within
the grasp of high quality, ground-based astrometric studies as well as the Gaia mission.

A remarkable aspect of the Sgr system is that the Sun presently lies within a kiloparsec
of the Sgr debris plane (MSWO).  The orbital pole of the plane, 
$(l_p,b_p)=(272, -12)^{\circ}$, means that the line of nodes of its intersection with the 
Galactic plane is almost precisely coincident with the Galactic $X_{GC}$ axis.  Thus, 
the motions of Sgr stars {\it within} this plane are almost entirely contained in their 
Galactic $U$ and $W$
velocity components, whereas the $V$ motions of stars in the Sgr tidal tails are almost entirely 
a reflection of the {\it solar motion} (see Fig.\ 1).  To the degree that the $V$
distribution of stars is not completely flat in Figure 1 is due to the slight amount of streaming
motion projected onto the $V$ motions from the 2$^{\circ}$ Sgr orbital plane tilt from $X_{GC}$, compounded by (1) Keplerian variations in the space velocity of stars as a function of orbital 
phase, as well as (2) 
precessional effects that lead to $\Lambda_{\sun}$-variable departures of
Sgr debris from the nominal best fit plane to all of the debris. 
As shown in Johnston et al. (2005), the latter is negligible for trailing debris but is much larger for the leading debris, which is on average closer to the Galactic center and
dynamically older compared to the trailing debris when viewed near the Galactic poles.
In addition, because the leading debris gets arbitrarily close to the Sun (\S2.3), projection
effects make it more complicated to use for the present purposes.  Additional problems
with the leading debris are discussed in Law et al. (2005), Johnston et al. (2005) and 
in \S4.   In contrast, the
Sgr trailing tail is beautifully positioned in a fairly equidistant arc around the Sun
for a substantial fraction of its stretch across the Southern Galactic hemisphere (MSWO).
This arcing band of stars situated almost directly ``beneath" the Sun along the $X_{GC}$
line provides a remarkable zero-point reference against which to make a direct 
measurement of the solar motion
{\it almost completely independent of the distance to the Galactic center. }

Because of this particular configuration of the trailing arm debris, almost all of the reflex
solar motion is contained in the $\mu_l \cos(b)$ component of the proper motion at parts
of the stream away from the Galactic pole.  Working in the observational, 
proper motion regime means
that vagaries in the derivation of {\it individual} star distances can be removed from the
problem, as long as the system is modeled with a proper {\it mean} distance for the
Sgr stream as a function of $\Lambda_{\sun}$.
Figure 2a shows three general regimes of the trailing arm $\mu_l \cos(b)$ trend:
(1) $\Lambda_{\sun}  \gtrsim100^{\circ}$ where $\mu_l \cos(b)$ is positive and roughly
constant, (2) the region from $100^{\circ} \gtrsim \Lambda_{\sun} \gtrsim 60^{\circ}$
 where $\mu_l \cos(b)$ flips sign as the debris
passes through the South Galactic Pole to shift the Galactic longitudes of the trailing arm
by $\sim180^{\circ}$,
and (3) 
$\Lambda_{\sun} \lesssim 60^{\circ}$, where $\mu_l \cos(b)$ is negative 
and becomes smaller with decreasing $\Lambda_{\sun}$ (because the Sgr stream becomes
increasingly farther).  The sign flip in the $\mu_l \cos(b)$ trend is a useful happenstance
in the case where
one has proper motion data that is not tied to an absolute reference frame but which is at 
least robust to systematic zonal errors: In this case one need only measure the 
peak to peak amplitude of $\mu_l \cos(b)$ for the trailing arm stars to obtain (two times) the 
reflex motion of the Sun.

\begin{figure}[!ht]
\plotfiddle{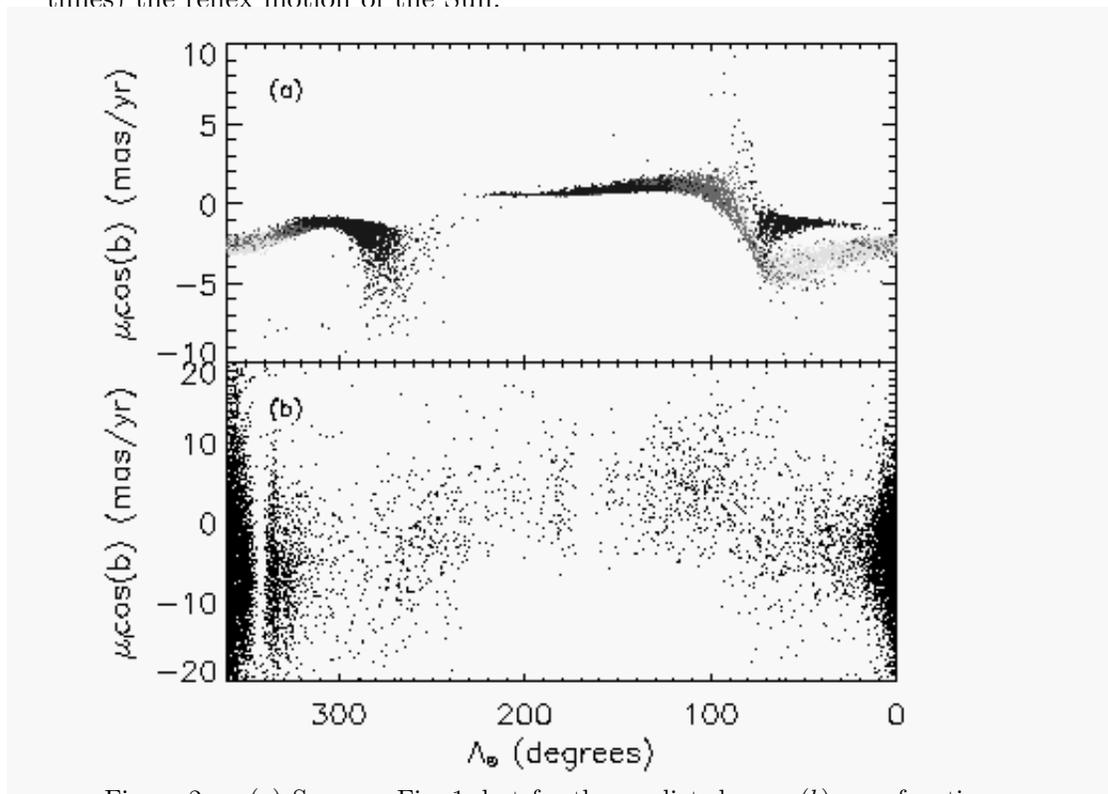}{90truemm}{0}{80}{80}{-239}{-185}
\caption{(a) Same as Fig.\ 1, but for the 
predicted $\mu_l \cos(b)$ as a function of orbital longitude.
(b) Observed UCAC2 $\mu_l \cos(b)$ for 2MASS M giants in Galactic regions dominated
by Sgr stream stars (all M giants within 7 kpc of the nominal Sgr plane and having
heliocentric distances of 15-30 kpc).  Note the differing vertical scales of the two panels.
}
\end{figure}

The intrinsic radial velocity dispersion of the Sgr trailing arm has been measured to be
$\sim$10 km s$^{-1}$ (Majewski et al. 2004a); assuming symmetry in the two 
transverse dimensions of the stream gives an intrinsic proper motion dispersion of the Sgr
trailing arm of $\sim$0.1 mas yr$^{-1}$ (see Fig.\ 2a).  Thus, until SIM-quality proper motions
exist, 
the measurement of the reflex solar motion by this method will be dominated by the error in 
the proper motions.  For example,
a measure of the peak-to-peak variation in Figure 2a good to a few percent
(a solar velocity good to $\sim$5 km s$^{-1}$) 
would require approximately 200 stars (say, $V \sim 15$ M giants) 
near the longitude of 
each Figure 2a trailing arm
peak measured to about 2 mas yr$^{-1}$ precision with no zonal systematics.
Present astrometric catalogs are just short of being able to do this:
The Hipparcos catalog does not go faint enough, the Southern Proper Motion Survey 
(Girard et al. 2004) has not yet covered sufficient amounts of the appropriate areas on 
the sky,
and the second USNO CCD Astrograph Catalog (UCAC2; Zacharias et al. 2004) has several 
times larger random errors than this as well as comparably-sized
zonal systematic errors at the appropriate magnitudes 
(N. Zacharias, private communication).  However, 
as a demonstration of how only modest advances
in all-sky proper motion precisions are needed to make a definitive measurement, 
Figure 2 shows a direct comparison of the $\mu_l \cos(b)$ trend for Sgr 
M giants using UCAC2 proper motions for 2MASS M giants.  Impressively, the overall expected
$\mu_l \cos(b)$ trends can be seen, but the large scatter and systematic shifts in the trailing arm
motions belie the limits of UCAC2 accuracies at $V \sim 15$.  Even a factor of two 
improvement in the UCAC2 random errors {\it and} elimination of zonal errors would likely
lead to a definitive measurement of the solar motion.  It is not unreasonable to expect
advances in all-sky proper motion catalogues at this level within a decade, but in any case SIM
will {\it easily} obtain the necessary proper motions (and parallaxes) of selected 
Sgr trailing arm giants.

\subsection{Sagittarius in the Solar Neighborhood}

The proximity of the Sun to the Sgr orbital plane has another interesting consequence:
The best fitting model to the Sgr debris (Law et al. 2005) predicts that the Sgr leading arm
penetrates the Galactic plane quite near the 
Sun.  For only 2\% of its orbit around the Galactic center is the Sun 
closer to the Sgr orbital plane on the side of the Galaxy where the leading arm is expected
to cross the stellar Galactic disk.  The observed width of the stream in other places (MSWO)
suggests, then,
that members of the Sgr stream should lie quite close to the Sun as they rain down from
the direction of the North Galactic Cap.  Extrapolation from the observed M 
giant density using the 47 Tuc-like luminosity function of Silvestri et al. (1998) finds that there 
should be at least one Sgr star within 30 pc of the Sun.  
The problem is to identify these  nearby stars
among the much higher density of stars from the Milky Way. 
Even the Sgr M giants --- an unusual stellar type in dSph galaxies --- are swamped by M giants
from the Galactic thin and thick disks.  Distinctive chemical abundance patterns in Sgr stars
(McWilliam \& Smecker-Hane 2004) means these stars would stand out spectroscopically,
but their relative density to the disk is so low that an alternative means is needed to isolate 
these needles in a haystack.

The expected large downward motion of the Sgr leading arm means that constituent stars
(away from the Galactic poles) should have a negative $b$-component to their proper motion.  Unfortunately, contamination by
disk stars is worst at low latitudes, and to find a statistically significant signal of 
negative $\mu_b$ stars requires a compromise to searching at mid-latitudes.  Nevertheless,
a signal like that expected can be found.  Figure 3 shows the distribution of $\mu_b$ for
M giants 
(with $J-K_s > 0.9$) having  $|b| < 60^{\circ}$ but estimated 
$(X_{GC}^2+Y_{GC}^2)^{1/2} < 4$ kpc (to keep the M giants relatively near the Sun)
and $|Z_{GC}| > 2.5$ kpc (to remove the majority of disk M giants).  An excess
of M giants with negative $\mu_b$ can be seen, particularly 
in the second Galactic quadrant, but the size of the M giant
excess greatly exceeds that expected for the presence of {\it only} 
Sgr debris near the Sun.  Full phase space data will help determine 
whether multiple halo substructures contribute to the excess, but evidence for at least 
{\it two} nearby,
{\it downward}-moving halo groups was previously noted by Majewski et al. (1996).

\begin{figure}[!ht]
\plotfiddle{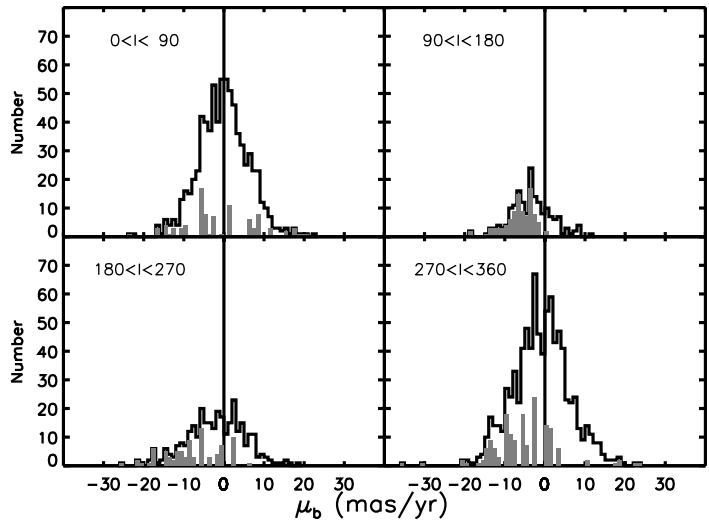}{70truemm}{0}{60}{60}{-179}{-204}
\vskip -5mm
\caption{Distribution by Galactic quadrant 
of $\mu_b$ for 2MASS-selected M giants with $J-K_s > 0.9$,
$|b| < 60^{\circ}$, $(X_{GC}^2+Y_{GC}^2)^{1/2} < 4$ kpc and $|Z_{GC}| > 2.5$ kpc.  The 
proper motions come from UCAC2, and are fairly reliable for these bright stars 
($V \lesssim 14 $).  Shaded histograms represent excesses of stars for a given 
$\mu_b$ bin compared to its counterpart, $-\mu_b$ bin.
The variation
in the total numbers of stars per Galactic quadrant reflects the incompleteness of UCAC2
at northern declinations.  In fact, the majority of 
stars shown for the second quadrant are in the southern Galactic hemisphere.
}
\end{figure}

\section{The RPMD as a Tool for Finding Halo Streams}

Figure 3 suggests that there is an imbalance of M giants with $\mu_b < 0$ near the Sun.
Are any of these excess stars moving fast enough to be from the Sgr leading arm?  Without
knowledge of the true metallicity of the M giants, ascertaining accurate distances to 
convert their proper motions to space velocities is a problem.
In such circumstances, the reduced proper motion diagram (RPMD; Luyten 1922) is
an extremely useful tool (Majewski 1999).  

A recently-developed, highly effective method to find halo substructure
is the identification of associated 
main sequence turn-offs (MSTOs) in ``field" color-magnitude
diagrams (``CMDs"; 
e.g., Newberg et al. 2002, Mart{\'{\i}}nez-Delgado et al. 2002, Ibata et al. 2003,
Majewski et al. 2004b).  This method works well to identify distant tidal debris structures
for which the MSTOs are well-defined and coherent; however,
the contrast of tidal stream features in the CMD lessens
as the distance, $r$, to a stream becomes comparable 
to its width, $\Delta r$.  In this case, the population ridge line spreads in magnitude by
$\Delta m = 5 \log{(1+\Delta r/r)}$ (see Fig.\ 4).  However, across the $\Delta r$ width of a stream, 
the transverse velocities, $V_{trans}$, of the constituent stars are roughly equal and
the stellar distances are then directly proportional to $ V_{trans}/\mu$.  Thus,
reduced proper
motions, $H_m = m + 5 + 5\log{\mu} = M + 5\log{V_{trans}}-3.38$, will yield coherent
structures when plotted versus color when simple stellar magnitudes will not.  

This advantageous property of the RPMD is especially useful 
for the study of nearby coherent streaming
where the range of distances for member stars can range by large factors,
as may be the case for the Sgr leading arm.
Figure 5 shows the RPMD of the M giant sample in the upper right panel of Figure 3,
separated by stars with positive and negative $\mu_b$. 
The $(J-K_s, K_s)$ ridge line for [Fe/H] = 0.0  and -0.6 giants (Ivanov \& Borissova 2002),
shifted to $V_{trans} = 40$ and 350 km s$^{-1}$ 
respectively, are shown to represent the expected loci of Milky Way disk-like and Sgr-like stars.  
As may be seen, the excess of nearby M giants with $\mu_b < 0$
seems to include a population lying near the locus of ``Sgr-like" 
reduced proper motion.  
Figures 3 and 5 suggest the presence of halo substructure, some consistent with Sgr debris,
in the solar neighborhood.  Follow-up spectroscopy on these stars for space velocities 
is in progress.

\begin{figure}
\plotfiddle{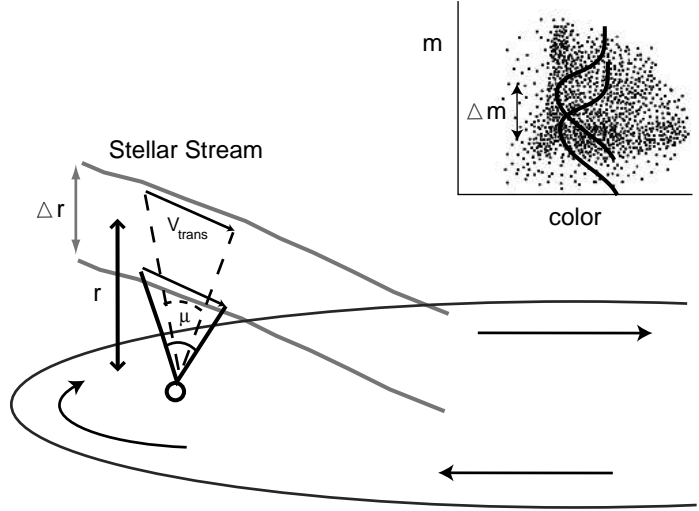}{62truemm}{0}{51}{51}{-170}{-81}
\vskip -5mm
\caption{Cartoon showing Sun ({\it circle}) in rotating Galactic disk with a 
tidal stream passing nearby.  Magnitude spreading in the CMD occurs when a stream's
mean distance approaches its width, $r \rightarrow \Delta r$, but
a well-defined isochrone can be recovered by considering
instead the reduced proper motion, $H_m = m + 5 + 5\log{\mu}$. }
\vskip -6mm
\end{figure}

\begin{figure}
\plotfiddle{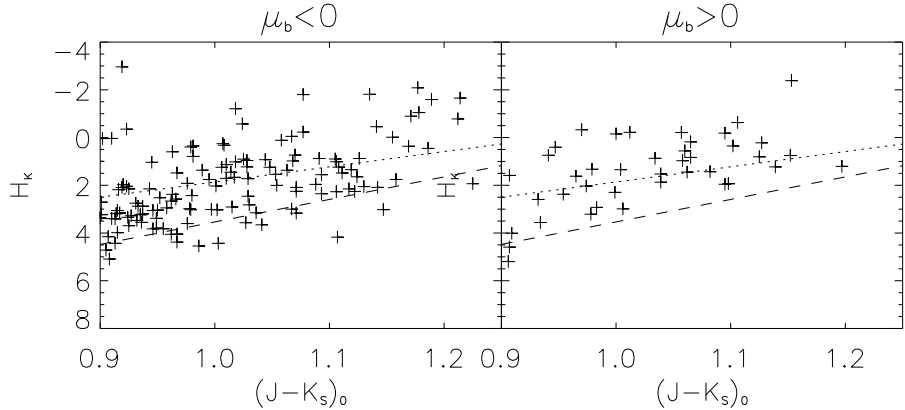}{50truemm}{0}{75}{75}{-220}{-370}
\vskip -5mm
\caption{RPMD for the stars in quadrant 2 from Fig.\  3, with stars having $\mu_b<0$
on the left and those with $\mu_b>0$ on the right .  The ridge lines correspond
to giant stars with [Fe/H] = 0 and $V_{trans} = 40$ km s$^{-1}$ (disk-like M giants; {\it dotted lines})
and with [Fe/H] = -0.6 and $V_{trans} = 350$ km s$^{-1}$ (Sgr-like M giants; {\it dashed lines}).
}
\vskip -6mm
\end{figure}

The RPMD has also recently been used to verify the reality of more distant M giants associated with the Triangulum-Andromeda structure (Rocha-Pinto et al. 2004) and
the Monoceros-Argus stellar stream (Rocha-Pinto et al. 2005).
This is an especially
effective technique to apply to these particular halo substructures, because they lie at low
Galactic latitude where the late dwarf foreground is extremely dense.  
Even with use of $JHK_s$ photometry to distinguish M giant from M dwarf stars, a small
``leakage" of nearby dwarf contaminants can swamp the giant star signal of these very diffuse
halo substructures, and, in the case of TriAnd, the system is sufficiently metal-poor
that it has few M giants to begin with.  On the other hand, in the RPMD, the TriAnd 
(G and K spectral type) giant branch can still be discerned
(Fig.\ 6).

\begin{figure}
\plotfiddle{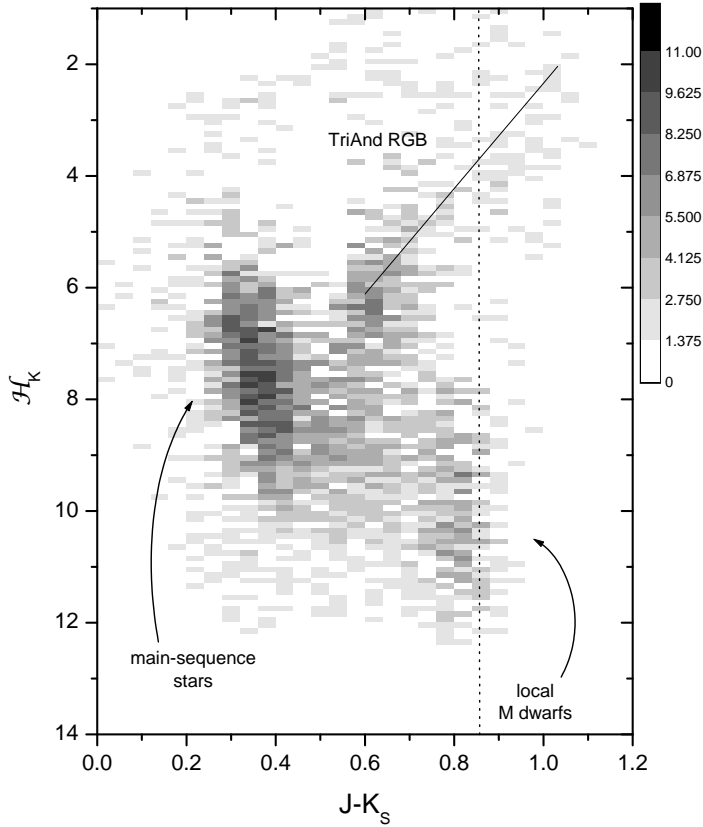}{96truemm}{0}{50}{50}{-144}{-30}
\vskip -5mm
\caption{Hess-like near infrared RPMD (with greyscale as square root of bin density)
for 2MASS stars with $11.0 < K_s < 12.5$ in the 
Triangulum-Andromeda region.  
See Figure 3 of Rocha-Pinto et al. (2004) for
more details.}
\vskip -6mm
\end{figure}

\section{Measuring the Milky Way Potential and Lumpiness from Streams}

Extended tidal streams 
are useful probes of the shape of the dark halo of the Milky Way, much as polar rings
have been used to measure the shapes of extragalactic halos (e.g., Sackett et al. 1994).  
However, results to date using the Sgr stream have been controversial.  Initially, the 
apparent coplanarity of dozens of carbon stars thought to be associated with the Sgr 
stream was taken as evidence that the Galactic halo is spherical (Ibata et al. 2001), and 
the results for M giants in the partially released 2MASS database seemed to confirm that 
(Ibata et al. 2002a).  On the other hand, Mart{\'{\i}}nez-Delgado et al. (2004) argued that the 
extant data on the Sgr stream prior to MSWO was more consistent with a flatter, 
$q=0.85$ halo.  With full 2MASS sky coverage, Law et al. (2004) and 
Johnston et al. (2005) have shown that 
only slight precession in the Sgr leading and trailing arms as traced by M giants suggests 
that the halo is only slightly flattened, with $q=0.92$ the formally best fitting halo potential.  

Recently, however, Helmi (2004) has argued for a {\it prolate} halo, with $q\sim1.25$!
Such a model alleviates a discrepancy between the observed leading Sgr arm 
M giant radial velocities (see Law et al. 2004, 2005) and much higher radial velocities predicted 
for these stars in spherical or oblate Galactic halos.  In prolate halo models, the Sgr  
leading arm {\it passes over} the position of the Sun by as much as 50 kpc at its 
highest Galactic latitudes and intersects the Galactic disk well outside the solar circle; in this
configuration, the space motions of the leading arm project less significantly into the radial velocities (and more into the proper motions).  
In contrast, with $q \le1$ potentials the Sgr leading arm is more directly pointed at us, 
the leading arm intersects the disk closer to the position of the Sun, and
the large space velocities of the stream stars are projected more along the line of sight.  
Unfortunately, the total observational picture appears contradictory: Low velocities in
leading arm stars support the prolate model, whereas the apparent presence of Sgr stars in
near the Sun (\S2.3) supports spherical/oblate halo models.  But an
additional problem with prolate halo models 
is that they precess the Sgr debris in the {\it opposite} direction to that which
is observed; indeed, based on the measured precession, 
prolate models with $q\ge1.25$ are ruled out at the 5-$\sigma$ level (Johnston et al. 2005).    
Interestingly, if the Milky Way is either only slightly oblate {\it or} prolate it is apparently an unusual
galaxy by the standards of either cosmological numerical simulations with cold dark matter
(CDM; e.g., Dubinski \& Carlberg 1991, Jing \& Suto 2002, Bullock 2002) 
or observations of extragalactic systems (e.g., 
Sackett et al. 1994, Sackett 1999,
Buote et al. 2002, Hoekstra et al. 2004), for which more oblate ($q \sim 0.5-0.7$) halos are typical and for which significant non-alignment between angular momentum 
and structural minor axes is apparently rare (e.g., Bailin \& Steinmetz 2004).

But if the weight of evidence points to a slightly oblate model, 
we are still left with the thorny problem of explaining the unexpectedly low leading arm 
radial velocities.   A solution proposed by Johnston et al. (2005) is that there may have been evolution of the orbit of Sgr, so that the leading arm debris has been perturbed from the 
nominal predictions of models like those shown in Figure 1.  Some evidence for evolution of
the Sgr orbit has recently been presented by Pakzad et al.
(2004).  Clearly, proper motion data for stars all along the Sgr leading arm is a priority, because 
the various halo models are also discriminated by how Sgr debris wraps within its orbital plane; because the latter is roughly the Galactic $X_{GC}-Z_{GC}$ plane, the information 
is contained in the trends of $U$ and $W$ with $\Lambda_{\odot}$ (see, e.g., Fig.\ 1).
Extensive, systematic astrometric study of substantial numbers of Sgr tidal tail stars will enable 
even more complex models of the Milky Way-Sgr interaction to be constrained.  

In this vein we note that, based on starcount models,
Newberg \& Yanny (this proceedings) argue that the spheroid of the Milky Way
may in fact be {\it triaxial}.  To test such a potential will require additional streams in 
different orientations.  With SIM we plan to measure proper motions of tidal debris stars 
separated by at least tens of degrees from their parent satellites as a means to
map the distribution of mass in the Galaxy.  With accuracies of tens of $\mu$as yr$^{-1}$
for of order 100 tidal debris stars per tidal stream the potential enclosed with each stream orbit should be estimable to $\sim$10\%.  With of order ten streams 
having different radii and orientations, we can determine radial trends in the potential and
test triaxiality.  Additional halo tidal streams aside from Sgr have already been found 
(e.g., Fig.\ 6) as a result of vigorously active search campaigns.

Because they are intrinsically dynamically cold, tidal streams are also extremely sensitive
to the {\it lumpiness} of the Galactic halo.  Large lumps gravitationally
``heat" cold streams (e.g., Ibata et al. 2002b, Johnston, Spergel \& Haydn 2002).  However, while
lumpy halos are a common property of CDM models, measurements of the radial velocity dispersion along the trailing Sgr arm (Majewski et al. 2004a) reveal at most
only slight heating over the past few Gyr --- results most consistent with a rather smooth
halo (although some ``lucky" configurations of lumpier halos cannot be ruled out with these 
data).  Tighter constraints on halo lumpiness would come from repeating measurements
on debris streams born of intrinsically colder (i.e. lower mass) progenitors, like
those from globular clusters.  In principle it is possible to measure proper motion
dispersions in the tidal tails, but here the astrometric requirements are demanding:  
To measure the transverse velocities to 3 km s$^{-1}$ accuracy (sufficient to measure
velocity {\it dispersions} to an order of magnitude better accuracy for samples of
only dozens of stars) for
a stream at 60 kpc requires proper motion accuracies of 10 $\mu$as yr$^{-1}$ for $V \sim 18$
K giant members.  However, SIM will make such measurements possible within a decade.

\acknowledgements 
We appreciate funding by NASA/JPL through the Taking Measure of the Milky Way
Key Project for SIM PlanetQuest, the National Science Foundation, the
David and Lucile Packard Foundation, and the F.H. Levinson Fund of the
Peninsular Community Foundation.



\end{document}